\documentclass[useAMS,usenatbib]{mn2e}
\usepackage{graphicx}
\usepackage{amssymb}
\title[Disc wind in the HH 30 binary models]{Disc wind in the HH 30 binary models}
\author[L. V. Tambovtseva and V. P. Grinin]{L. V. Tambovtseva$^{1}$\thanks{E-mail:
tamb@gao.spb.ru (LVT); grinin@gao.spb.ru (VPG)} and V. P.
Grinin$^{1,2}$\\
$^{1}$Main Astronomical Observatory Pulkovo, Pulkovskoe shosse 65,
St. Petersburg 196140, Russia\\
$^{2}$The Astronomical Institute of the St. Petersburg University,
Petrodvorets, St. Petersburg, 198904, Russia}

\begin{document}

\date{Accepted    Received ; }

\pagerange{\pageref{firstpage}--\pageref{lastpage}} \pubyear{XXXX}

\maketitle

\label{firstpage}

\begin{abstract}
Recent interferometric observations of the young stellar
object(YSO) HH 30 have revealed a low velocity outflow in the
$^{12}$CO J=1-2 molecule line (Pety et al. 2006). We present here
two models of the low velocity disc winds with the aim of
investigating an origin of this molecular outflow. Following
Andlada et al. (2006) we treated HH 30 as a binary system. Two
cases have been considered: i) the orbital period $P$ = 53 yrs and
ii) $P \le$ 1 yr. Calculations showed that in the first case the
outflow cone had a spiral-like structure due to summing the
velocities of the orbital motion and the disc wind. Such a
structure contradicts the observations. In the second case, the
outflow cone demonstrates a symmetry relatively to the system axis
and agrees well with the observations.
\end{abstract}


\section{Introduction}
The Herbig--Haro object HH 30 is a well--known YSO with a flared
circumstellar (CS) disc seen practically edge--on (Burrows et al.
1996; Ray et al. 1996) and highly collimated optical jets
propagating perpendicular to the disc, firstly observed by Mundt
\& Frid (1983). The star itself is not visible, it is detected
only through radiation scattered by the CS disc. According to Pety
et al. (2006) (hereafter P06), HH 30 is a star with a mass of
0.45$M_\odot$, spectral type around of M1 and age of (1 - 4)$10^6$
years. The millimeter observations showed that it is surrounded
with the CS Keplerian disc with the radius of about 420 AU and the
mass of about $4 \cdot 10^{-3} M_\odot$. Optical observations
showed the disk of about 450 AU in diameter (Burrows et al. 1996).

Since 1996 this object has been a subject of intense
investigations by different authors (Bacciotti et al. 1999;
Stapelfeldt et al. 1999; Wood et al. 2000, 2002; Cotera et al.
2001; Watson \& Stapelfeldt 2004). Optical observations with the
\textit{Hubble Space Telescope} (\textit{HST}) as well as the
millimeter interferometry revealed a strong variability and
asymmetry in the radiation coming from the CS disc. Burrows et al.
(1996) detected changes in the brightness ratio of the upper and
lower nebulae between 1994 and 1995, and the presence of a weak
lateral brightness asymmetry. Stapelfeldt et al. (1999) found that
in 1998 the right (north--northwest) side of the upper reflection
nebula brightened dramatically, and the other side faded relative
to 1995. The system had appeared nearly symmetrical about the jet
axis in 1995, but in 1998 one side of the disc was four times
brighter than the other.

The interpretation of the HH 30 brightness variability and asymmetry has a great importance.
Stapelfeldt et al. (1999) excluded large-scale motions of the matter in the outer disc as a possible
reason of such variations. They argued that the outer disc acts as a screen, on which moving
illumination patterns are projected from the inner disc or the central star. They suggested two
mechanisms which are able to form such light changes: (i) illumination by the star with bright
accretion hot spots (Wood \& Whitney (1998)) and (ii) presence of voids or clumps in the inner disc
casting beams or shadows onto the outer disc. Each of these mechanisms can produce asymmetry in the
nebulae brightness, which shifts from one to another side of the disc. Watson \& Stapelfeldt (2004,
2007) did not find the period to be correlated with the stellar rotation period or any other period.
According to them, the variability of HH 30 has the more complex character and caused by the changes
in the CS extinction in the inner disc. Tambovtseva and Grinin (2008) showed that such extinction
variations can be caused by the dust component of the azimuthally non-homogeneous disc wind. A
knowledge of the asymmetry mechanism would give a possibility to determine in detail a structure of
the nearest regions of the young star that has not been done with the help of the telescopes. Also
this permits us to study more definitely the physical processes leading to formation of such
structures.

One of the main result of the HH 30 study obtained during recent years, was the detection of the slow
matter outflow in the $^{12}$CO(2-1) molecule line by P06. This outflow was highly asymmetric
(one-sided) in spite of the presence of relatively symmetric jets, and originated from the inner parts
of the disk, had a conical morphology with a 30$^\circ$ half opening angle and a radial velocity of
about 12 km s$^{-1}$. P06 concluded that the observed CO emission arose from material that has been
directly launched from the disc and moved ballistically in the first few 100 AU from the star; the
re-collimation could occur at the distances larger than 1000 AU. According to their interpretation,
the high-velocity optical jet and low-velocity outflow are associated with different parts of the disc
wind. The former corresponds to its densest part ($< 1 AU)$ and the latter to its outer part (5 - 15
AU). The mass loss rate of the outflow $\dot M_w$ was estimated as $6.3\times 10^{-8} M_\odot$
yr$^{-1}$; it is almost 60 times more than that of the jet equaled to $\approx 10^{-9} M_\odot$
yr$^{-1}$. Because however of their velocity differences it follows that a majority of the kinetic
energy of the disc wind ($L_w \approx 6\cdot 10^{31}$ erg c$^{-1}$) is contained in the narrow
collimated jet, while the bulk of the mass loss occurs via the low-velocity wind component.

Recently Anglada et al. (2007) revealed periodic changes in the jet's trajectory and connected this
with the binarity of HH 30. They concluded that the structure of the HH 30 jet can be described by a
wiggling ballistic jet, arising either by the orbital motion of the jet source (i.e., the secondary)
around the primary or by precession of the jet axis because of the tidal effects of the secondary. In
the first scenario, the orbital period is 53 years, and the sum of masses is 0.25 - 2 $M_\odot$. In
the second scenario, the mass of the jet source (i.e., the primary) has to be $ \sim 0.1 - 1 M_\odot$,
the orbital period is less than 1 year, and the mass of the companion less than a few $\times 0.01
M_\odot$. In both scenarios the separation of the binary does not exceed 18 AU. Knowing the size of
the flared disc the authors conclude that this is circumbinary rather than circumstellar disc.

In this paper we model the disc wind in the binary system in the framework of the two scenarios
suggested by Anglada et al. (2007). In section 2 we briefly formulate the problem for the first
scenario and present results of calculations, in section 3 we do the same for the precession scenario.
Then we discus results and make the conclusion.

\section{Disc wind in the young binary system}
We consider the wind from the low--mass secondary companion moving in a circular orbit in the young
binary system. The companion accretes matter from the remnants of the protostellar cloud, the
so-called circumbinary (CB) disc. In detail the model is presented in our early paper (Grinin \&
Tambovtseva (2002)). Here is a brief description of the model. Following Artymowicz \& Lubow (1996),
we treat the low--mass companion as the main accretor.  The wind particles, consisting of gas and dust
in the standard proportion, are ejected from the surface of its accretion disc within a cone having
fixed inner and outer angles. It is supposed that in the frame of the secondary, the outflow has an
azimuthal symmetry. In the coordinate system of the primary, the wind becomes asymmetric due to the
vector summing the particles and the orbital motion velocities. Trajectories of the wind particles in
the gravitational field of the primary have been calculated in the ballistic approach. We imitated the
quasi--continuous process by ejecting the particles in a small and equal time interval. Computation
stopped when the particles reached the outer radius $R_{out}$ or the equatorial plane of the binary.

In order to reproduce the molecular outflow in HH 30 we used parameters estimated by P06: the outflow
matter is mainly located on the thin walls of a cone with an semi--opening angle of $30^\circ$. The
radial velocity $V_w = 12 \pm$ 2 km/s, the tangential velocity component is absent. The mass loss rate
$\dot M_w = 6.3 \times 10^{-8} M_\odot$ yr$^{-1}$.

Taking this into account, we adopt the following parameters for our model: mass of the primary $M_* =
0.45M_\odot$, and the eccentricity $e = 0$. In order to provide the orbital period of 53 yrs, the
radius of the orbit $r_0 = 11$ AU has been chosen. The initial radial velocity of the wind particle,
in units of the corresponding Keplerian velocity, was an input parameter and varied from 2 to 5. The
Keplerian velocity at $r_0 = 11$ AU is about 6 km s$^{-1}$. Three models of the outflow are presented
in Fig. 1, and their parameters given in the Table (models A, B, C).

\begin{figure}
\centering
\includegraphics[width=90mm]{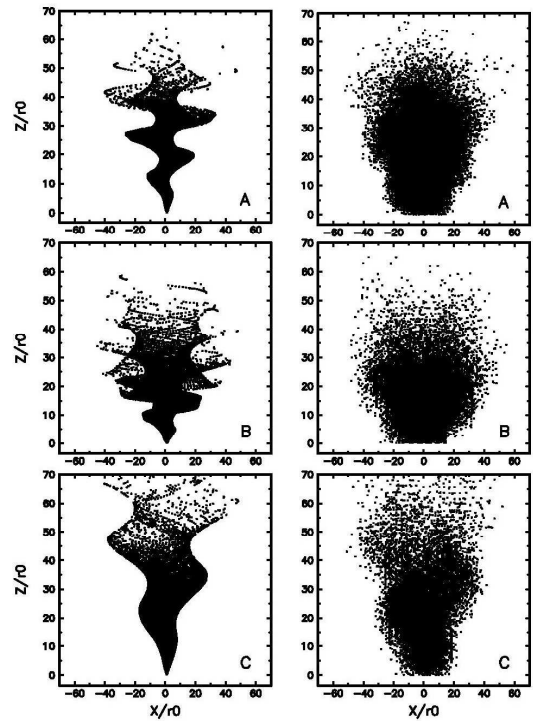}\\
\caption{The distribution of the wind particles in the common (with the primary) envelope created by
the disc wind of the secondary component in the models A, B and C (the XZ projection). The left panel:
the computed particles distribution; the right panel: the same after smoothing procedure (details are
in the text). Letters refer to the model. Distances X and Z are given in units of the orbital radius.}
\end{figure}

\begin{table}
\caption[]{Parameters of the models}
\begin{flushleft}
\begin{tabular}{cccc}
\hline
  Model & $\theta$ & V &  U\\
\hline
  A & 30 & 3 & 0\\
  B & 45 & 3 & 0\\
  C & 30 & 5 & 0\\
  D & 30 - 40 & 2 & 1 \\
\hline
\end{tabular}
\end{flushleft}
\end{table}

Thus, in the models considered below, the common parameters are
the parameters of the orbit and a zero tangential component of the
wind velocity $U$. The models differ with radial velocities of the
wind particles $V$ and angles of their ejection $\theta$ (Table
1). The angles are counted from the symmetry axis. Figure 1 (the
left panel) displays the computed distribution of the wind
particles in the common envelope in projection on the XZ plane.
The right panel of Fig. 1 gives the same distribution after the
smoothing procedure, which has been made as follows. We set a
"theoretical beam" with a spatial resolution equaled to that used
by P06, when observing the emission in the $^{12}$CO(2-1)
molecule. Then, using this beam we scanned a theoretical image
shown on the left side of Fig. 1. We show only upper planes in the
figure.

An increase in the particle ejection angle (models A and B) leads to more amorphous structures, while
an increase in the initial radial velocity of the wind (models A and C) to the larger step of the
spiral, but a common structure persists. The radial velocities estimated from the observations do not
permit us to increase the start velocities infinitely. We checked also the other variants (low radial
velocities, random ejection angles), but a cone structure has not been obtained in the framework of
this scenario.

\begin{figure}
\includegraphics[width=8cm]{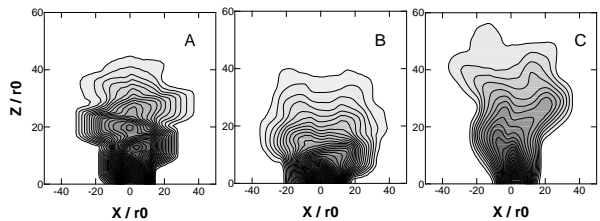}\\
\caption{Isodensity contours for models A, B and C. The details are in the text.}
\end{figure}

If we assume, as has been done by P06, that the emissivity of the gas is proportional to the local
density, and the brightness distribution is proportional to the column density, one may consider the
isodensity map as an intensity map. Such maps are given in Fig. 2 for models A, B and C. Contours of
equal densities were calculated using the smoothed distributions of the prob particles. The contours
are shown in the grey scale where the outermost curve corresponds to the level of 100 probe particles,
and the maximal level is about 3000 - 4000 particles depending on the model. Each level differs from
the next one of 100 probe particles. The method of the transition from the probe particles to the real
column densities (in cm$^{-2}$) is described in our paper (Grinin \& Tambovtseva 2002).

Comparison of the isophotes obtained for these models with the observed ones (Fig. 1 in the paper by
P06) leads to the conclusion that this scenario is not realized in the HH 30 system.

\section{Precession scenario}
Anglada et al. (2007) suggested a second scenario: the wiggling jet structure is caused by precession
of the jet axis due to tidal perturbations from the low mass companion. In this case the binary system
is very close. The jet source is an accretion disc of the primary and the question arises: where does
the low velocity wind come from? P06 concluded that for a single star, the source of the biconical
molecular outflow lies on the surface of the accretion disc at 5 - 15 AU from the center. This means,
that in the close binary with the low--mass companion, located at the distance of $\lesssim 1$ AU from
the central star, the inner disc of the primary cannot be a source of the molecular outflow, because
the launched region would be near the inner boundary of the CB--disc. In the case of a circular orbit,
one can assume that the inner boundary of the CB--disc possess an axial symmetry. Therefore, the
initial conditions for the disc wind will not depend on the azimuth, and, thus, an outflow produced by
such a wind will be axially symmetric. In this scenario, the wind particles simultaneously start from
the inner CB--disc in small time intervals. As above, calculations have been made using a ballistic
approach. Trajectories of the particles have been calculated exactly as in the first case, differing
only in initial and boundary conditions, and with that the orbital motion of companions did not
influence anyhow the particles kinematics. The radius of the orbit $r_0$ = 0.75 AU. With such a radius
and the mass of the star 0.45 $M_\odot$, the period of the system is about 1 year. According to
calculations by Artymowicz \& Lubow (1994) the inner radius of the CB--disc $r_{CB}$ is 3-4 times
greater than the orbital radius $r_0$. We adopted $r_ {CB}=4r_0 \approx 3 AU$. In this case the
Keplerian velocity at the inner radius of the CB - disc is about 13 km/s.

\begin{figure}
\includegraphics[width=6cm]{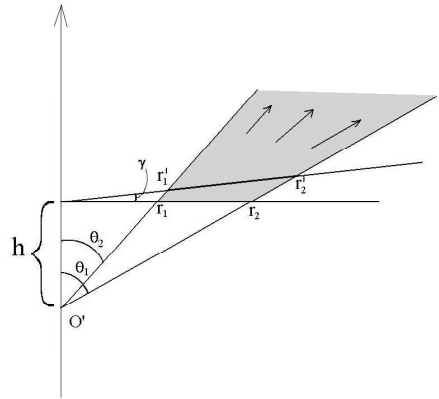}\\
\caption{The wind geometry.}
\end{figure}

Table 1 includes parameters of the model D considered in the second scenario. The radial and
tangential components of the initial velocity of the wind particles are expressed in units of the
Keplerian velocity at $r_{CB}$. The particles are ejected in the cone constrained with the angles
$\theta_1 = 40^\circ$ and $\theta_2 = 30^\circ$ (Fig 3). As calculation show, in this case the disc
wind produces axially symmetric outflow with parameters close to those observed. The isodensity map
calculated for such an outflow is presented in Fig. 4.

Let us estimate the optical depth of the disc wind $\tau$ in this
model of HH 30. Denote $\Delta\theta = \theta_1 - \theta_2$.
Taking into account, that the thickness of the molecular outflow
cone is small (P06), we assume that $\theta_1 \approx \theta_2
\equiv \theta$. Then,  $\Delta\theta \ll \theta$. With the help of
the Eq. (1) from the paper by Grinin \& Tambovtseva (2006), one
can obtain the optical depth of the cone "wall" along the radius
$r_1'r_2'$, which at the small $\gamma$ can be replaced by the
radius $r_1r_2$ which is parallel to the disc plane and intersects
the disc axis at the distance $h$ from the point O$'$ (Fig. 3):
\begin{equation} \tau = \frac{\dot M_w}{V_w}\,\frac{\kappa_\lambda}{2\pi h
\sin{\theta}}\,.
\end{equation}
Here $\dot{M}_w$ and $V_w$ are the mass loss rate and the velocity of the disc wind material
respectively, $\kappa_\lambda$ is a coefficient of absorbtion by the dust component of the disc wind
at the wavelength $\lambda$. It should be noted, that when deriving this formula we took into
consideration, that the outflow in HH 30 occurs with a strong deviation from mirror symmetry: the
major part of the matter flows in the upper cone.

\begin{figure}
\centering
\includegraphics[width=5cm]{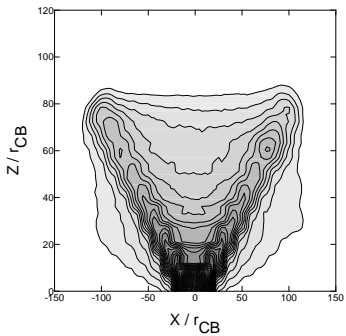}\\
\caption{The same as in Fig. 2 for the model D.}
\end{figure}

Substituting the values of $\dot{M_w}$ and $V_w$ in the expression (1) and adopting $\theta =
30^\circ$ and $\kappa_\lambda$ = 250 cm$^2$/g (that corresponds to the wavelength $\approx$ 5000 \AA
\hspace{0.1cm} for the CS dust), we obtain the optical depth in the base of the wind (h = 5 AU) $\tau
\approx 3$. Thus, in the model of the outflow adopted in this section, the optical depth for the base
turns out to be of the order of few units. We estimated the angle $\gamma$ with the help of the scale
height law for HH 30 (see Table 3 given by P06). We obtain that the peripheral parts of the flared
disc are illuminated by the stellar radiation under the angle $\gamma = 10 - 15$ degrees. This means,
that at the higher regions (the radius $r_1'r_2'$ in Fig. 3), the disc wind could still be opaque for
radiation from the star. If the wind has an azimuthally non-homogeneous structure, this may result in
the irregular illumination of the disc, and thus in the variability of the radiation scattered by the
disc.

We also checked the effects of such parameters as radial and tangential velocity components on the
wind density distribution, and did not present these results since all of them were analogous to that
discussed above. An increase in the initial radial velocity gives a more extended outflow, the greater
tangential component of the initial velocity leads to the greater opening angle, but a principal
structure remains the same: there is a concentration of the wind particles towards the cone's walls.

\section{Discussion}
Let us put aside the problem of the binarity of HH 30 and discuss other possible mechanisms for the
formation of the molecular outflow. A comprehensive overview of the molecular outflow models is
presented by Arce et al. (2007) (see also the references therein). All of them can be separated into
four broad classes: (1) wind-driven shells, (2) jet-driven bow shocks, (3) jet-driven turbulent flows,
and (4) circulation flows. In the first three models the molecular outflow is entrained by an
underlying wind or jet, in the fourth, it is formed by deflected infalling matter.

Pety et al. (2006) discuss the second entrainment scenario as a possible mechanism of the outflow
origin. According to this mechanism, the observed outflows consist in ambient molecular gas that has
been put into motion by large bow shocks propagating down the underlying jets (e.g. Gueth \&
Guilloteau 1999). This mechanism explains well outflows observed on large scales ($\gg$ 1000 AU). In
the case of HH 30, the outflowing molecular gas is continuously collimated down to the very close
vicinity of the star, on spatial scales that are smaller than the disk size. This implies that the
outflow arises from material that has been directly launched from the disk surface. Also, it is
difficult to understand why CO emission is absent in the southern lobe. In the framework of the
jet-driven flow models this means that properties of the ambient medium in the north and south
directions differ strongly . Thus, the main arguments against entrainment and pro disc wind scenario
are a conical geometry, a radial velocity near the launching region, and a lack of CO emission in the
south lobe. For these reasons, P06 ruled out the jet-driven flows from considerations in favor of disc
wind mechanism. \footnote{It should be noted that entrainment mechanism would lead to the conical
structure of the molecular outflow but arguments mentioned above prevent its application.}.

Nevertheless, if we assume that the molecular outflow originates from the outer part of the disk wind,
launched at a few AU from the star, another problem arises: how to explain the interval between the
jet and the outflow, in other words, the lack of detected material at intermediate velocities and
angles? The binary status of HH 30 helps to avoid this difficulty. In both models of the binary system
considered in the present paper, the jet and the disc wind originate from different sources. It should
also be taken into account that in the binary system with the low-mass companion a cavity free of
matter arises due to the gravitational perturbations (Artymowicz \& Lubow 1994), therefore, this pause
may be naturally explained.

Based on the analysis of the observed (Hartigan et al. 1995) and computed forbidden lines, Kwan \&
Tademaru (1995) conclude that a substantial fraction of the emitting wind particles has low
velocities. Can the low--velocity part of the disc wind, originating from the warm disc (Ferreira,
Dougados \& Cabrit 2006, Ferreira 2007), produce such a molecular outflow as observed in HH 30? This
question needs to be explored. In the case of a single star a discontinuity in launch regions of jets
and disc winds would be a more serious issue than in the case of the binary system, but the disc wind
problem in binaries remains to be investigated.

\section{Conclusion}

We calculated the trajectories of the disc wind motion using a
ballistic approach within the framework of the two binary system
scenarios for HH 30. In the first model, the disc wind originates
from the accretion disc of the secondary companion; in the second
model, the source of the low--velocity disc wind is the inner
region of the circumbinary disc.

Calculations showed that in the first case, the matter outflow had a spiral-like structure due to the
orbital motion of the wind source around the primary. Such a structure does not agree with the axially
symmetric cone morphology of the outflow in HH 30 observed by Pety et al. (2006). The cone outflow
geometry is typical for the second model, in which the well collimated and highly accelerated jet, and
the slow molecular outflow, originate from the different regions of the binary system: the jet from
the inner disc of the primary, and the molecular outflow from the inner boundary of the CB disc. The
dust component of the azimuthally non-homogeneous disc wind could be a source of the variable
extinction both in the case of HH 30, and other young stellar objects\footnote{Cotera et al. (2007)
studying the images of the young stellar objects obtained with \textit{HST}, found three other objects
with the variable lateral asymmetries analogous to that detected in HH 30.}.

\section*{Acknowledgments}
We thank the referee, Tom Ray, for helpful comments and suggestions which improved the clarity of the
paper. This work was performed as part of the "Origin and Evolution of Stars and Galaxies" Program of
the Presidium of the Russian Academy of Sciences under support of INTAS grant no. 03-51-6311 and grant
no. NSh-8542.2006.2.

\end{document}